\documentclass[reprint,nofootinbib,amsmath,amssymb,aps,prl]{revtex4-1}
\usepackage{graphicx}
\usepackage{bm}
\usepackage{hyperref}
\def\vec#1{\bm{#1}}
\def\abs#1{\left\lvert#1\right\rvert}
\def\vev#1{\langle#1\rangle}

\let\Im\relax\DeclareMathOperator{\Im}{Im}
\let\Ei\relax\DeclareMathOperator{\Ei}{Ei}
\let\erf\relax\DeclareMathOperator{\erf}{erf}
\def\eb{\varepsilon_B}
\def\ef{\varepsilon_F}

\begin{document}

\title{Bulk viscosity and contact correlations in attractive Fermi gases}
\author{Tilman Enss}
\affiliation{Institut f\"ur Theoretische Physik,
  Universit\"at Heidelberg, D-69120 Heidelberg, Germany}
\date{\today}

\begin{abstract}
  The bulk viscosity determines dissipation during hydrodynamic
  expansion.  It vanishes in scale invariant fluids, while a nonzero
  value quantifies the deviation from scale invariance.  For the
  dilute Fermi gas the bulk viscosity is given exactly by the
  correlation function of the contact density of local pairs.  As a
  consequence, scale invariance is broken purely by pair fluctuations.
  These fluctuations give rise also to logarithmic terms in the bulk
  viscosity of the high-temperature nondegenerate gas.  For the
  quantum degenerate regime I report numerical Luttinger-Ward results
  for the contact correlator and the dynamical bulk viscosity
  throughout the BEC-BCS crossover.  The ratio of bulk to shear
  viscosity $\zeta/\eta$ is found to exceed the kinetic theory
  prediction in the quantum degenerate regime.  Near the superfluid
  phase transition the bulk viscosity is enhanced by critical
  fluctuations and has observable effects on dissipative heating,
  expansion dynamics and sound attenuation.
\end{abstract}
\maketitle

The bulk viscosity is a fundamental transport property which
determines friction and dissipation in fluids during hydrodynamic
expansion \cite{landauVI, forster1975}.  In particular, scale
invariant fluids can expand isotropically without dissipation and
therefore have zero bulk viscosity \cite{son2007}.  In a generic
interacting fluid, instead, a nonzero value of the bulk viscosity
quantifies the breaking of scale invariance in physical systems
ranging from QCD \cite{karsch2008, moore2008, romatschke2009,
  akamatsu2018} to condensed matter \cite{taylor2010, enss2011,
  taylor2012, dusling2013, chafin2013scale, elliott2014observation,
  fujii2018}.  An intriguing example is the two-dimensional dilute
Fermi gas, where the classical model is scale invariant but a quantum
scale anomaly breaks this symmetry \cite{pitaevskii1997, olshanii2010,
  hofmann2012, vogt2012}; this has recently been observed via
breathing dynamics in cold-atom experiments \cite{holten2018,
  peppler2018, murthy2019}.

The bulk viscosity is necessary to understand and predict the
real-time evolution and hydrodynamic modes of dissipative quantum
fluids and to quantitatively interpret current experiments.  However,
measurements of the bulk viscosity remain challenging even for
classical fluids \cite{dukhin2009}.  Now a novel experimental probe
via the dissipative heating rate due to a change in scattering length
has been proposed for atomic gases \cite{fujii2018}.  It is therefore
important to compute the bulk viscosity theoretically for quantum
gases, which moreover includes predictions for the classical gas in
the high-temperature limit.

The bulk viscosity is defined as the correlation function of local
pressure (the trace of the stress tensor).  Since it vanishes in a
scale invariant system, only the scale breaking part of the pressure
contributes, the so-called trace anomaly \cite{martinez2017,
  czajka2017, fujii2018}.  This provides a formal link between the
breaking of scale invariance and bulk viscosity.  The bulk viscosity
of the nonrelativistic, strongly interacting Fermi gas has been
calculated from kinetic theory in the nondegenerate high-temperature
limit \cite{dusling2013, chafin2013scale} and in the low-temperature
superfluid state \cite{escobedo2009, hou2013}.  Its value is largest
in the strongly coupled region of the BEC-BCS crossover
\cite{zwerger2012} near unitarity, but not precisely at unitarity
where is must vanish by scale invariance \cite{werner2006unitary,
  son2007, enss2011}.  Furthermore, hydrodynamic fluctuations give
rise to nonanalytic corrections to the bulk viscosity at small
frequencies \cite{onuki2002, martinez2017}.  However, key open
questions include the bulk viscosity in degenerate Fermi gases at
strong interaction, the relative importance of bulk and shear
viscosity, and critical scaling near the superfluid phase transition.

In this work, I rewrite the bulk viscosity of the dilute Fermi gas as
a correlation function of the contact density of local fermion pairs.
This exact mapping explicitly links the bulk viscosity to pairing
fluctuations as the relevant degrees of freedom and provides a genuine
strong-coupling formulation which is valid in the whole BEC-BCS
crossover including the quantum critical regime \cite{nikolic2007,
  sachdev2011, enss2012crit}.  New results include (a) dominant
logarithmic corrections to the bulk viscosity at high temperature, (b)
numerical Luttinger-Ward results for the quantum degenerate gas
throughout the BEC-BCS crossover predict a large bulk viscosity well
observable with current experimental technology, (c) the transport
ratio of bulk to shear viscosity deviates from the kinetic theory
prediction in the quantum degenerate regime, and (d) critical scaling
near the superfluid transition is less singular than predicted
\cite{kadanoff1968, onuki2002}, but pairing fluctuations dynamically
enhance the scale anomaly.


\textit{Bulk viscosity.}---%
The bulk viscosity $\zeta$ is defined as the stress correlation function
\cite{kadanoff1963, landauIX, taylor2010}
\begin{align}
  \label{eq:zeta}
  \zeta(\omega) = -\frac1{\omega d^2} \Im \int_0^\infty dt\,
  e^{i\omega t} \int d^dx \vev{[\Hat\Pi_{ii}(\vec x,t),\Hat\Pi_{jj}(0,0)]},
\end{align}
where the trace of the stress tensor
$\Hat\Pi_{ii}(\vec x,t) = d\cdot \Hat P$ determines the pressure
operator $\Hat P$ in dimension $d$.  The two-component dilute Fermi
gas is described by the Hamiltonian density \cite{zwerger2012}
\begin{align}
  \label{eq:Ham}
  \Hat{\mathcal H} = \sum_\sigma \psi_\sigma^\dagger
  \bigl(-\frac{\nabla^2}{2m}\bigr) \psi_\sigma + g_0 
  \psi_\uparrow^\dagger \psi_\downarrow^\dagger \psi_\downarrow
  \psi_\uparrow .
\end{align}
The first term denotes the kinetic energy with fermion operators
$\psi_\sigma(\vec x,t)$.  The attractive contact interaction in the
second term is characterized by the $s$-wave scattering length $a$.
For a given value of $a$, the bare coupling strength $g_0$ is
determined according to $g_0^{-1} = -(m/2\pi)\ln(a\Lambda)$ in two
dimensions (2D) and $g_0^{-1} = (m/4\pi)(1/a-2\Lambda/\pi)$ in 3D,
with ultraviolet momentum cutoff $\Lambda$.  The trace of the stress
tensor is given by the scale variation of the Hamiltonian
\cite{hofmann2012},
\begin{align}
  \label{eq:Pi}
  d\cdot \Hat P
  = \Hat\Pi_{ii}
  = [\Hat{\mathcal H},i\Hat D]
  = 2\Hat{\mathcal H}+
    \begin{cases}
    \frac{\Hat{\mathcal C}}{2\pi m} & \text{(2D)},\\
    \frac{\Hat{\mathcal C}}{4\pi ma} & \text{(3D),}
  \end{cases}
\end{align}
where the dilatation operator
$\Hat D=\int d^dx\, \vec x\cdot m\vec j(\vec x)$ generates scale
transformations.  The first term on the right-hand side is the scale
invariant result $[\Hat{\mathcal H},i\Hat D]=2\Hat{\mathcal H}$.  If
only this is present, the pressure is proportional to the Hamiltonian
and commutes with itself in \eqref{eq:zeta}, hence the bulk viscosity
$\zeta(\omega)\equiv0$ vanishes identically in the scale invariant
case \cite{werner2006unitary, son2007, enss2011, zwerger2016}.

The second term, in turn, is proportional to the local pair contact
density
$\Hat{\mathcal C} = -m^2(\partial\Hat{\mathcal H}/\partial g_0^{-1}) =
m^2g_0^2\psi_\uparrow^\dagger \psi_\downarrow^\dagger \psi_\downarrow
\psi_\uparrow$ \cite{tan2008energetics}.  Scale invariance is
recovered for the ideal quantum gas where $\Hat{\mathcal C}=0$, and
also for the 3D unitary Fermi gas where $1/a=0$ at the scattering
resonance.  A nonzero bulk viscosity therefore quantifies the breaking
of scale invariance, which is generally expected in the interacting
Fermi gas, except at unitarity.

\textit{Contact correlation.}---%
By conservation of energy, the Hamiltonian in \eqref{eq:Pi} does not
contribute to the pressure commutator \eqref{eq:zeta}, and the bulk
viscosity is given by the correlator of the scale breaking term.  The
scaling violation in the trace of the stress tensor is the so-called
trace anomaly \cite{martinez2017, fujii2018}
\begin{align}
  \label{eq:traceano}
  \Hat\Pi_\text{an} \equiv \Hat\Pi_{ii}-2\Hat{\mathcal H}
  = c_d\,\Hat{\mathcal C},
\end{align}
where $c_d = -(\partial g_0^{-1}/\partial \ln\abs a)/m^2$ denotes the
scale variation of the bare coupling (beta function).  For the dilute
gas, the equilibrium bulk viscosity is thus exactly given by the contact
correlator,
\begin{align}
  \label{eq:zetaC}
  \zeta(\omega) = -\frac{c_d^2}{\omega d^2} \Im \int_0^\infty dt\,
  e^{i\omega t} \int d^dx \vev{[\Hat{\mathcal C}(\vec x,t),
  \Hat{\mathcal C}(0,0)]}.
\end{align}

The contact operator is the term in the Hamiltonian which couples to
the scattering length.  In linear response, the bulk viscosity thus
captures how the local pair contact density at time $t$ changes in
response to a variation of the scattering length at earlier time
$t=0$ \cite{fujii2018},
\begin{align}
  \label{eq:chi}
  \chi(\vec x,t) \equiv
  \vev{[\Hat{\mathcal C}(\vec x,t),\Hat{\mathcal C}(0,0)]} = 
  \begin{cases}
    2\pi m\bigl(\frac{\partial\vev{\Hat{\mathcal C}(\vec x,t)}}
    {\partial\ln a(0,0)}\bigr)_s &
    \text{(2D)}, \\
    -4\pi m\bigl(\frac{\partial\vev{\Hat{\mathcal C}(\vec x,t)}}
    {\partial a^{-1}(0,0)}\bigr)_s &
    \text{(3D)}
  \end{cases}
\end{align}
at constant entropy per particle $s=S/N$.  The time dependent contact
response captures how quickly the contact adjusts to a change in
scattering length; this directly determines the dynamical bulk
viscosity according to Eq.~\eqref{eq:zetaC}.  This makes the contact
correlation, and hence the dynamical bulk viscosity, directly
accessible in cold atom experiments where the scattering length can be
controlled in time by the magnetic field near a Feshbach resonance and
the time evolution of the contact has already been measured using RF
spectroscopy \cite{bardon2014, luciuk2017}.


\textit{Viscosity sum rule.}---%
Since the pressure operator is hermitean, the dynamical bulk viscosity
is an even and positive function of frequency, $\zeta(\omega)\geq0$
\cite{taylor2010}.  The integral over all frequencies in
Eqs.~\eqref{eq:zetaC}, \eqref{eq:chi} immediately yields the bulk
viscosity sum rule \cite{taylor2010, taylor2012} with $\mathcal C =
\vev{\Hat{\mathcal C}}$,
\begin{align}
  \label{eq:sumrule}
  S \equiv \frac2\pi \int_0^\infty d\omega\, \zeta(\omega) =
  \begin{cases}
    -\frac{1}{8\pi m}
    \bigl(\frac{\partial \mathcal C}{\partial \ln a}\bigr)_s &
    \text{(2D)}, \\
    \frac{1}{36\pi ma^2} \bigl(\frac{\partial \mathcal
        C}{\partial a^{-1}}\bigr)_s & \text{(3D)}.
  \end{cases}
\end{align}
Using the Tan adiabatic relation to express the contact
$\Pi_\text{an} = c_d\,\mathcal C = (\partial\mathcal E/\partial\ln\abs
a)_s$ as the scale variation of the energy density $\mathcal E$
\cite{tan2008energetics, hofmann2012, werner2012}, the sum rule is
given by the scale ``susceptibility''
$S = -(1/d^2)(\partial^2\mathcal E/\partial(\ln\abs a)^2)_s \geq0$ in
$d$ dimensions.  The sum rule is taken at constant entropy per
particle to ensure that the bulk viscosity of the ideal gas is zero
\cite{taylor2010}.

\textit{Pair fluctuations.---}%
The local contact density can equivalently be interpreted as the
density operator
$\Hat{\mathcal C} = \Hat \Delta^\dagger \Hat\Delta(\vec x)$ of the
local fermion pair field
$\Hat\Delta(\vec x) = mg_0\psi_\downarrow(\vec x) \psi_\uparrow(\vec
x)$.  The bulk viscosity thus depends directly, and only, on pairing
fluctuations within the attractive Fermi gas; it is given exactly by
the four-point pair correlation function
$\chi(\vec x,t)=\vev{[\Hat\Delta^\dagger \Hat\Delta(\vec x,t),
  \Hat\Delta^\dagger \Hat\Delta(0,0)]}$.  One can anticipate that the
bulk viscosity has a strong signature at the superfluid phase
transition which is driven by pair fluctuations (see below).  While
pair fluctuations are strong also at unitarity, the prefactor
$c_d^2\sim 1/a^2$ ensures that $\zeta$ vanishes in this case.

To summarize, the bulk viscosity is the response function of the trace
anomaly and is therefore sensitive to scaling violation.  For the
dilute quantum gas, the trace anomaly is proportional to the contact
density of local pairs and depends only on the pairing properties.
This establishes the link between pairing \cite{murthy2018} and the
quantum scale anomaly \cite{murthy2019} suggested by recent
experiments in 2D Fermi gases.


\textit{Analytical results.}---%
The contact correlations and bulk viscosity can be computed exactly in
several limiting cases: at (i) zero density (two-body), (ii) high
frequency, and (iii) high temperature (virial expansion).

The zero-density case (i) is determined solely by two-body physics.
In this limit, the only source of dissipation is the dissociation of a
bound molecule at the two-body binding energy $\eb=\hbar^2/ma^2$; this
yields a high-frequency tail above the threshold $\omega>\eb$ to break
a pair \cite{taylor2012,SM},
\begin{align}
  \label{eq:zeta02D}
  \zeta_\text{2D,vac}(\omega)
  = \frac{\mathcal C_0}{4m\omega}\,
  \frac{\Theta(\omega-\eb)}{\ln^2(\omega/\eb-1)+\pi^2}
\end{align}
in 2D, where $\mathcal C_0$ denotes the two-body contact.  In 3D, a
two-body bound state exists only on the BEC side for $a>0$, and
\begin{align}
  \label{eq:zeta03D}
  \zeta_\text{3D,vac}(\omega)
  = \frac{\mathcal C_0\,\Theta(a)}{36\pi ma}\,
  \frac{\sqrt{\eb(\omega-\eb)}\, \Theta(\omega-\eb)}{\omega^2}.
\end{align}
This two-body result serves to disentangle the dissipation due to
two-body pair breaking from the genuine many-body bulk viscosity below
\cite{taylor2012}.

In the limit (ii) of high frequency $\omega\gg\ef,T$, the contact
correlator is evaluated at small times where it factorizes as
$\chi(\vec x,t\to0) \simeq m^2\Gamma(\vec x,t) \mathcal C(0,0)$; at
large frequency, the pair propagator $\Gamma(\vec x,t)$ approaches the
zero-density form \cite{SM}.  It follows immediately that for
$\omega\to\infty$ the bulk viscosity is proportional to the contact
density and decays with a characteristic frequency dependence,
\begin{align}
  \label{eq:zetahigh}
  \zeta(\omega\to\infty) =
  \begin{cases}
    \frac{\mathcal C}{4m\omega\ln^2(\omega/\eb)} & \text{(2D)}, \\
    \frac{\mathcal C}{36\pi a^2(m\omega)^{3/2}} & \text{(3D)}.
  \end{cases}
\end{align}
This derivation reproduces earlier results \cite{hofmann2011,
  taylor2012, goldberger2012} in a dramatically simpler calculation.
The zero-density results \eqref{eq:zeta02D} and \eqref{eq:zeta03D}
approach the high-frequency limit with two-body contact density
$\mathcal C_0$.  However, the exact high-frequency limit is more
general and holds at arbitrary density, temperature and interaction in
terms of the total contact density $\mathcal C(n,T,a)$.  This
asymptotic behavior is important because it guarantees convergence of
the sum rule \eqref{eq:sumrule}.

\begin{figure}[t]
  \centering
  \includegraphics[width=\linewidth]{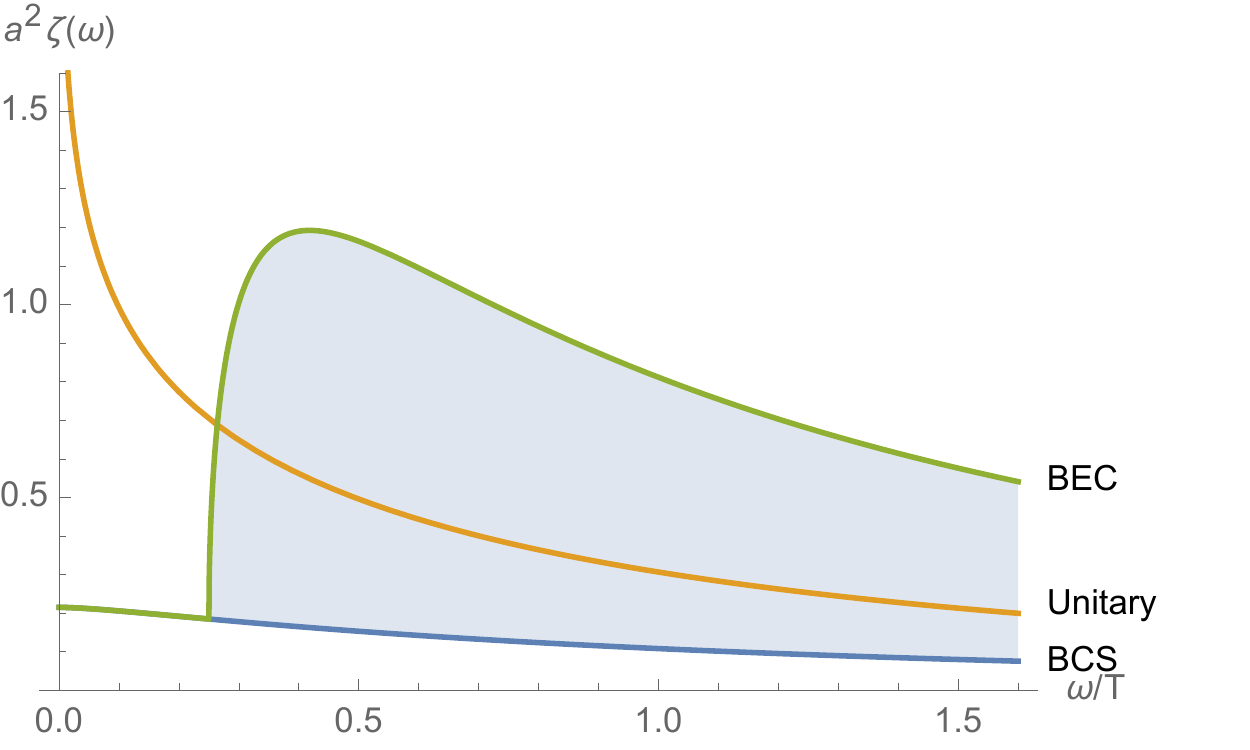}%
  \llap{\raisebox{2.78cm}{\includegraphics[width=.55\linewidth]{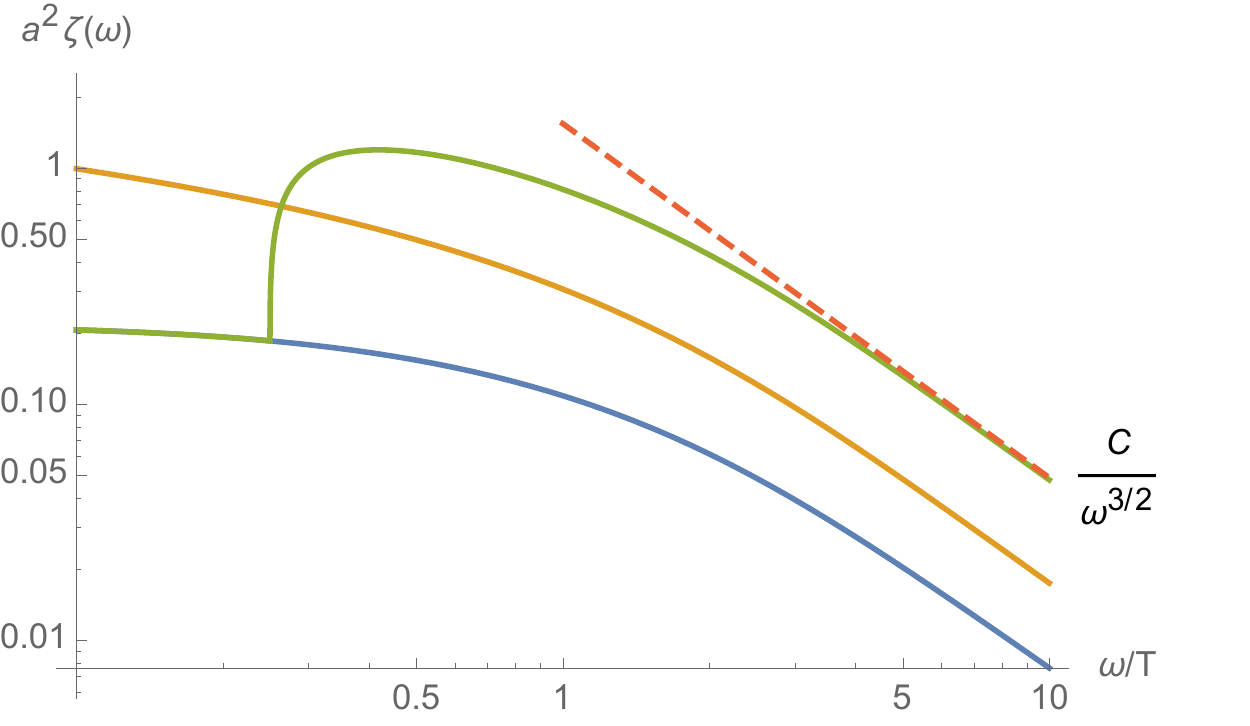}}}
  \caption{Dynamical bulk viscosity
    $\zeta(\omega)/[\sqrt2 z^2/9\pi a^2\lambda]$ vs frequency $\omega$
    in the high-temperature limit \eqref{eq:zeta3Dvir}.  From top to
    bottom: BEC ($v=0.5$, green), Unitary ($v=0$, orange), BCS
    ($v=-0.5$, blue).  Inset: logarithmic plot shows exact
    high-frequency asymptotics \eqref{eq:zetahigh}
    proportional to $\mathcal C/\omega^{3/2}$ (dashed).}
  \label{fig:virdynam}
\end{figure}
\begin{figure}[t]
  \centering
  \setbox1=\hbox{\includegraphics[width=\linewidth]{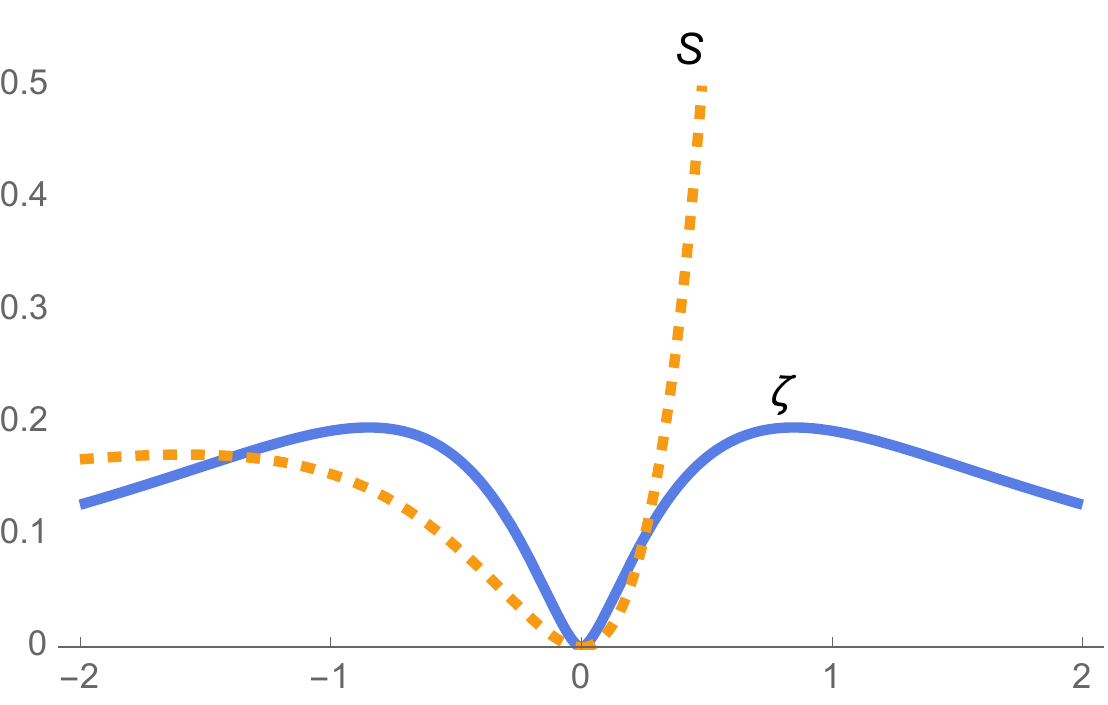}}%
  \includegraphics[width=\linewidth]{fig2a}%
  \llap{\makebox[\wd1][l]{\raisebox{2.5cm}%
      {\hskip6mm\includegraphics[width=.5\linewidth]{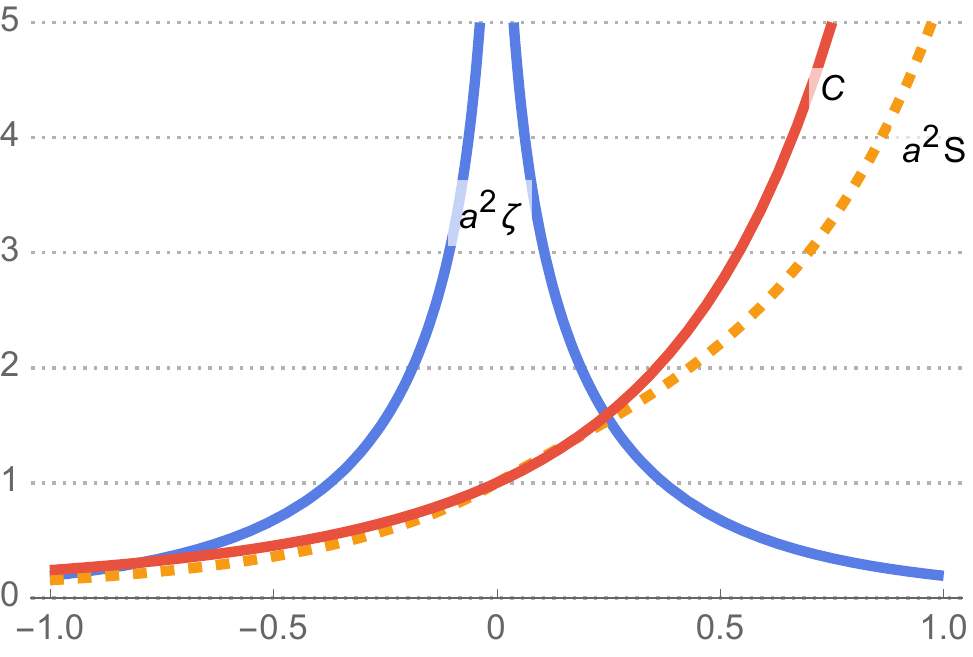}}}}
  \caption{Bulk viscosity $\zeta(v)$ vs interaction
    $v=\lambda/a\sqrt{2\pi}$ in the high-temperature limit.  Viscosity
    $\zeta/[2^{3/2}z^2/9\pi\lambda^3]$ \eqref{eq:zeta3Dvirdc} (solid
    blue), sum rule $S/[2^{3/2}z^2T/9\lambda^3]$ \eqref{eq:sum3Dvir}
    (dashed orange).  Inset: Contact correlation
    $a^2\zeta$ (solid blue), contact sum rule
    $a^2S$ (dashed orange) and contact
    $\mathcal C/[16\pi z^2\lambda^{-4}]$ (red).}
  \label{fig:virdc}
\end{figure}

Finally, the dynamical bulk viscosity can be computed exactly in the
high-temperature limit (iii) by virial expansion \cite{dusling2013,
  chafin2013scale}.  To second order in fugacity $z=e^{\beta\mu}$, the
pair distribution $b(\varepsilon) = z^2 e^{-\beta(\varepsilon+2\mu)}$
is combined with the zero-density spectral function to yield \cite{SM}
\begin{multline}
  \label{eq:zeta3Dvir}
  \zeta_\text{3D,vir}(\omega>0)
  = \frac{2\sqrt2}9 z^2\lambda^{-3}v^2
  \frac{1-e^{-\beta\omega}}{\beta\omega} \\
  \times \Bigl[ \Theta(v)
  2ve^{v^2}\frac{\sqrt{\beta\omega-v^2}\,\Theta(\beta\omega-v^2)}{\beta\omega}
  \\
  + \frac1\pi \int_0^\infty dy\,\frac{e^{-y}\sqrt{y(y+\beta\omega)}}
  {(y+v^2)(y+\beta\omega+v^2)} \Bigr].
\end{multline}
Here, $v=(\lambda/a)/\sqrt{2\pi}$ denotes the dimensionless
interaction parameter as the inverse scattering length in units of the
thermal length $\lambda=\sqrt{2\pi/mT}$.  The dynamical viscosity has
two terms as illustrated in Fig.~\ref{fig:virdynam}: the first,
bound-continuum contribution occurs only on the BEC side $v>0$ and
arises from breaking up bound states at high frequency
$\abs\omega>\eb$, which leads to strong damping as seen before in the
two-body limit \eqref{eq:zeta03D}.  The second term is the
continuum-continuum contribution of dissociated pairs, which extends
over all frequencies but has most of its spectral weight at small
frequencies $\omega\lesssim\eb$.  At this order there is no
bound-bound contribution because an ideal Bose gas of bound pairs is
scale invariant; corrections arise from atom-dimer scattering at order
$\mathcal O(z^3)$.  Both contributions in \eqref{eq:zeta3Dvir} are
necessary to exhaust the sum rule (cf.\ Fig.~\ref{fig:virdc})
\begin{multline}
  \label{eq:sum3Dvir}
  S_\text{3D,vir}
  = \frac{2\sqrt2}9 z^2T\lambda^{-3} v^2 \Bigl[
  (1+2v^2)e^{v^2}(1+\erf(v)) \\
  + \frac{2v}{\sqrt\pi} - \Theta(v)4v^2e^{v^2} \Bigr].
\end{multline}
This agrees with the adiabatic derivative \eqref{eq:sumrule} of the
contact \cite{yu2009, dusling2013}
$\mathcal C_\text{3D,vir} = 16\pi z^2\lambda^{-4} \bigl[ 1+\sqrt\pi\,
ve^{v^2}(1+\erf(v))\bigr]$.

\begin{figure*}[t]
  \centering
  \includegraphics[width=.34\linewidth]{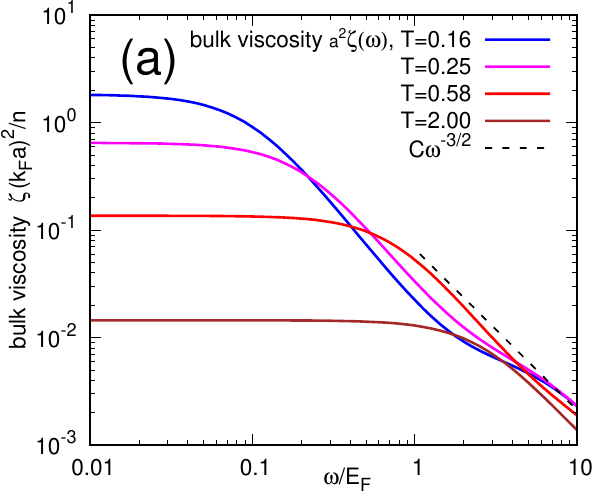}\hfill%
  \includegraphics[width=.327\linewidth]{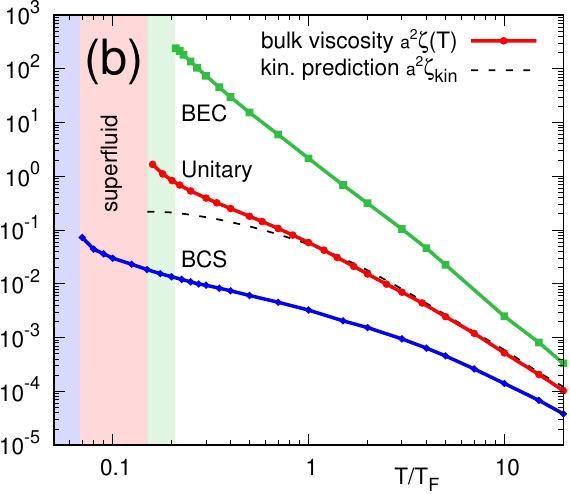}\hfill%
  \includegraphics[width=.327\linewidth]{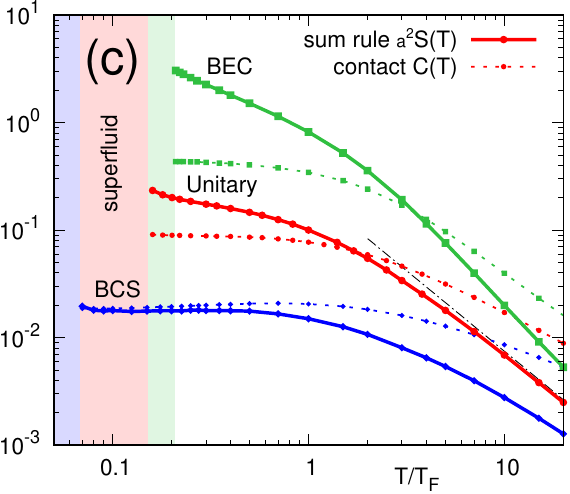}
  \caption{Bulk viscosity from Luttinger-Ward computation. (a)
    Dynamical bulk viscosity $\zeta(\omega)(k_Fa)^2/\hbar n$ vs
    frequency $\omega$ at unitarity, for increasing temperature from
    top to bottom (see legend); universal high-frequency tail
    $\zeta(\omega)\sim \mathcal C/\omega^{3/2}$ (dashed).
    (b) dc bulk
    viscosity $\zeta(T) (k_Fa)^2/\hbar n$ vs temperature for different interaction:
    from top to bottom BEC ($1/k_Fa=1$), Unitary ($1/k_Fa=0$), and BCS
    ($1/k_Fa=-1$).  Kinetic prediction $\zeta_\text{kin}(T) (k_Fa)^2/\hbar
    n$ \eqref{eq:zetakin} from shear viscosity (dashed).
    (c) Sum rule $S(T) (k_Fa)^2/n\ef$ (solid) and contact
    $\mathcal C(T)/k_F^4$ (dashed) vs temperature for different
    interaction; virial limit $T^{-3/2}$ (dot-dash).}
  \label{fig:lwdyn}
\end{figure*}

At unitarity $v\to0$, the analytical dynamical viscosity
\begin{align}
  \label{eq:zeta3Dviruni}
  \zeta_\text{3D,vir}^\text{unitary}(\omega)
  = \frac{2\sqrt2}{9\pi} z^2\lambda^{-3}v^2
  \frac{\sinh(\beta\omega/2)}{\beta\omega/2} K_0(\beta\omega/2).
\end{align}
At this order, the unitary contact correlation has a logarithmic
singularity $a^2\zeta\sim\ln(T/\omega)z^2$ for small frequencies from
the modified Bessel function $K_0(\beta\omega/2)$, as shown in
Fig.~\ref{fig:virdynam}.  The logarithmic singularity for small
frequencies corresponds via Fourier transform to the logarithmic
singularity of the bulk viscosity at long times,
$\zeta(t)\sim \ln(t)/(a^2t)$ \cite{maki2019}.  Precisely at unitarity,
the bulk viscosity vanishes for all frequencies due to the
$v^2\propto a^{-2}$ factor.  Throughout the BEC-BCS crossover, the dc
bulk viscosity is then given by (see Fig.~\ref{fig:virdc})
\begin{align}
  \label{eq:zeta3Dvirdc}
  \zeta_\text{3D,vir}(v)
  = \frac{2\sqrt2}{9\pi} z^2\lambda^{-3}v^2
  \bigl[-1-(1+v^2)e^{v^2}\Ei(-v^2) \bigr].
\end{align}
The exponential integral $\Ei(x)$ yields a logarithmic singularity in
scattering length $a^2\zeta \sim \ln(a^2/\lambda^2)z^2$ shown in the
inset of Fig.~\ref{fig:virdc}.  The singular coefficient of the virial
expansion is regularized by higher-order terms $\mathcal O(z^3)$ from
the fermionic self-energy \cite{enss2011, dusling2013}; these are
resummed in the Luttinger-Ward computation and yield a finite dc limit
in Fig.~\ref{fig:lwdyn}(b) below.

In 2D, there is always a bound state with binding energy $\eb>0$ even
for arbitrarily weak attractive interaction.  The dynamical bulk
viscosity is obtained as \cite{SM}
\begin{multline*}
  \zeta_\text{2D,vir}(\omega)
  = 2\pi z^2\lambda^{-2} \frac{1-e^{-\beta\omega}}{\beta\omega} \Bigl[
  \frac{\beta\eb e^{\beta\eb}\; \Theta(\omega-\eb)} {\ln^2(\omega/\eb-1)+\pi^2} \\
  + \int_0^\infty dy\, \frac{e^{-y}} {[\ln^2(yT/\eb)+\pi^2]
  [\ln^2((yT+\omega)/\eb)+\pi^2]} \Bigr].
\end{multline*}
The dc bulk viscosity is then approximately given by
\begin{align}
  \label{eq:zeta2Dvir}
  \zeta_\text{2D,vir}(\eb/T)
  \simeq \frac{2\pi z^2\lambda^{-2}}{[\ln^2(T/2\eb)+\pi^2]^2}.
\end{align}
This result for the bulk viscosity based on contact correlations is
similar in structure to the fermionic Boltzmann calculation
\cite{chafin2013scale} but larger by a factor $4\pi^2$, which is
necessary to satisfy the sum rule \cite{SM} and the high-frequency
asymptotics with the contact density \cite{ngampruetikorn2013,
  barth2014}
$ \mathcal C_\text{2D,vir} = 16\pi^2z^2\lambda^{-4}\bigl[ \beta\eb
e^{\beta\eb} + \int_0^\infty dy\, \frac{e^{-y}}{\ln^2(yT/\eb)+\pi^2}
\bigr].$


\textit{Luttinger-Ward results.}---%
The Luttinger-Ward (LW) technique is a diagrammatic strong-coupling
approach to fermions in the BEC-BCS crossover \cite{haussmann2007,
  bauer2014} which treats fermions $\psi_\sigma$ and the pair field
$\Delta$ on equal footing.  Its predictions for the unitary shear
viscosity \cite{enss2011} agree well with recent data
\cite{bluhm2017}, and similarly for spin diffusion \cite{enss2012spin,
  enss2019spin}.  In this work, I extend the previous LW approach to
compute the bulk viscosity \eqref{eq:zetaC} via the contact
correlation function \eqref{eq:chi}.  It uses the self-consistent pair
propagator $\Gamma$ and includes vertex corrections which represent
the scattering between pairs, resummed to arbitrary order \cite{SM}.
While contact vertex corrections are subleading in the
high-temperature limit and could be neglected, they are crucial in the
quantum degenerate regime and need to be included for an accurate
numerical solution.

The \emph{dynamical} bulk viscosity $\zeta(\omega)$ determines the
dissipation when the scattering length in Eq.~\eqref{eq:chi} is
modulated at frequency $\omega$; the hydrodynamic limit is obtained
for $\omega\to0$.  While $\zeta(\omega)$ vanishes at unitarity as
$1/a^2$, the contact correlations $a^2\zeta(\omega)$ are nonzero at
unitarity as shown in Fig.~\ref{fig:lwdyn}(a).  At low temperature
there is a pronounced peak at low frequencies $\omega\lesssim T$ that
crosses over into the universal high-frequency tail
$\zeta(\omega)\sim \mathcal C\omega^{-3/2}$ (dashed) for
$\omega\gtrsim\ef$.  At higher $T\gtrsim T_F$ the thermal peak for
$\omega\lesssim T$ leads directly into the tail.  The peak width
$\propto T$ is consistent with quantum critical scaling.

The dc bulk viscosity $a^2\zeta(T)$ shown in Fig.~\ref{fig:lwdyn}(b)
is one of the central results: it is largest near the superfluid
transition and decreases toward high temperature where pair
fluctuations become weaker, as discussed below.

\textit{Bulk/shear ratio.---}At high temperature kinetic theory
predicts the ratio of bulk viscosity $\zeta$ to shear viscosity
$\eta$,
\begin{align}
  \label{eq:zetakin}
  \zeta/\eta \propto (\Pi_\text{an}/\Pi_{ii})^2
  = [(P-2\mathcal E/3)/P]^2,
\end{align}
to be proportional to the squared pressure deviation from scale
invariance \cite{bluhm2012, dusling2013}.  Using LW bulk/shear and
thermodynamic data \cite{enss2011}, this is tested by comparing
$\zeta$ to the kinetic theory prediction
$\zeta_\text{kin}\equiv\eta[(P-2\mathcal E/3)/P]^2$, which is shown in
Fig.~\ref{fig:lwdyn}(b) as the dashed line.  There is very good
agreement with unit proportionality factor at high temperature
$T\geq T_F$, where a quasiparticle picture is expected to hold.
Consequently, the shear viscosity at high temperature is fully
determined by scale breaking pair fluctuations as reflected in $\zeta$
and in the contact.

In the quantum degenerate regime, the bulk viscosity grows
monotonously as the temperature is lowered toward the superfluid phase
transition and can reach large values $\zeta\gtrsim \hbar n$ near
$T_c$.  At low temperature, $\zeta>\zeta_\text{kin}$ and also
$\zeta/\eta>1$ can exceed unity since pair fluctuations near the
superfluid phase transition affect the bulk viscosity more strongly
than the shear viscosity.

\textit{Critical pair fluctuations.---}%
The fact that the bulk viscosity is the dynamical correlator of
order-parameter fluctuations $\Delta(\vec x,t)$ suggests that $\zeta$
might diverge at $T_c$ \cite{onuki2002}; instead, vertex corrections
in the LW calculation substantially reduce the contact vertex at low
momenta and render the bulk viscosity large but finite \cite{SM}.  The
absence of divergent critical scaling might depend on how the critical
point is approached, as found in QCD \cite{martinez2019}.

Finally, Fig.~\ref{fig:lwdyn}(c) shows the viscosity sum rule $S$.  It
is large in the quantum degenerate regime and decreases toward high
temperature as $T^{-3/2}$ \eqref{eq:sum3Dvir} (dot-dashed),
i.e., faster than the contact $\mathcal C\sim T^{-1}$ itself
(dashed) \cite{yu2009, enss2011, mukherjee2019}.


To conclude, the bulk viscosity identifies the breaking of scale
invariance with the strength of pair fluctuations, which become very
large near $T_c$ and on the BEC side.  This provides a strong
signature in cold atom experiments, either directly in the response of
the contact \cite{bardon2014, luciuk2017, fujii2018} to a change in
scattering length, or by modulating the scattering length periodically
and measuring the dissipative heating rate
$\dot{\mathcal E} = d^2a^2\zeta\cdot(\partial_t a^{-1})^2$
\cite{fujii2018} proportional to $a^2\zeta(\omega,T)$ shown in
Fig.~\ref{fig:lwdyn}(a,b), which is nonzero also at unitarity.
Further signatures of enhanced dissipation $\zeta$ can be found in the
hydrodynamic description of scaling or breathing dynamics
\cite{vogt2012, taylor2012, elliott2014observation, murthy2019} and
sound attenuation
$D_s = [\frac43\eta+\zeta+\kappa(c_v^{-1}-c_p^{-1})]/mn$
\cite{forster1975, patel2019}.

\begin{acknowledgments}
  \textit{Note added.} After submission, two other calculations
  \cite{nishida2019, hofmann2019} of the bulk viscosity in the
  high-temperature limit appeared, which agree with our results where
  applicable.

  \textit{Acknowledgments.} I acknowledge fruitful discussions with
  M.~Bluhm, G.~Bruun, J.~Hofmann, M.~Horikoshi, S.~Jochim, Y.~Nishida,
  J.~Pawlowski, T.~Sch\"afer, J.~Thywissen, W.~Zwerger, and
  M.\,W.~Zwierlein.  This work is supported by Deutsche
  Forschungsgemeinschaft (DFG) via Collaborative Research Centre
  ``SFB1225'' (ISOQUANT) and under Germany’s Excellence Strategy
  ``EXC-2181/1-390900948'' (Heidelberg STRUCTURES Excellence Cluster).
\end{acknowledgments}


\bibliography{all}


\appendix
\section{Supplemental material}

\subsection{Contact correlations}
The contact correlation can be written in terms of the pair field
$\Delta(\vec x,\tau)$ in imaginary time $\tau$ as
\begin{align}
  \label{eq:chitau}
  \chi(\vec x,\tau)
  = \vev{\Delta^\dagger\Delta(\vec x,\tau)
  \;\Delta^\dagger\Delta(0,0)}
\end{align}
with imaginary time ordering understood.  The pair propagator
$\vev{\Delta(\vec x,\tau) \Delta^\dagger(0,0)} = m^2 \Gamma(\vec
x,\tau)$ can be expressed in terms of the T matrix $\Gamma(\vec
x,\tau)$, and one can write the contact correlation as the scale
variation of the T matrix,
\begin{align}
  \label{eq:chicorr}
  \chi(\vec x,\tau)
  & = \frac{m^2}{c_d} \frac{\delta \Gamma(\vec x,\tau)}{\delta\ln \abs{a(0,0)}} \\
  & = m^4 \Gamma(\vec x,\tau) \Gamma(-\vec x,-\tau) + \text{vertex corrections}.\notag
\end{align}
In imaginary Matsubara frequency $i\omega_m$, the spatially integrated
contact correlator is given by
\begin{multline}
  \label{eq:chifreq}
  X(i\omega_m)
  = \int d^dx\, \chi(\vec x,i\omega_m) \\
  = \frac{m^4}{\beta V} \sum_{\vec q\varepsilon_n}
  \Gamma(\vec q,i\varepsilon_n)
  \Gamma(\vec q,i\varepsilon_n+i\omega_m)
  + \text{vtx.\ corr.},
\end{multline}
while the contact density itself is given by
$\mathcal C = (m^2/\beta V) \sum_{\vec q\varepsilon_n} \Gamma(\vec
q,i\varepsilon_n)$.  The vertex corrections are important in the
quantum degenerate case and are computed below using the
Luttinger-Ward approach.  At high temperature or low density, the
vertex corrections are subleading and the first term can be computed
analytically.  After analytical continuation to real frequency, the
retarded correlator reads
\begin{multline}
  \label{eq:Xret}
  X^R(\omega) = \frac{m^4}{V} \sum_{\vec q} \int d\varepsilon\,
  A(\vec q,\varepsilon) b(\varepsilon) \\
  \times [\Gamma^R(\vec
  q,\varepsilon+\omega) + \Gamma^A(\vec q,\varepsilon-\omega)]
  + \text{vtx.\ corr.}
\end{multline}
in terms of the retarded/advanced pair propagators and the pair
spectral function
$A(\vec q,\varepsilon) = -(1/\pi)\Im \Gamma^R(\vec q,\varepsilon)$.
This determines the dynamical bulk viscosity as the contact correlator
spectral function
\begin{multline}
  \label{eq:zetaX}
  \frac{d^2}{c_d^2}\,\zeta(\omega)
  = -\frac{\Im X^R(\omega)}{\omega} \\
  = \frac{\pi m^4}{\omega V} \sum_q \int_{-\infty}^\infty d\varepsilon\,
  A(\vec q,\varepsilon) A(\vec q,\varepsilon+\omega)
  [b(\varepsilon) - b(\varepsilon+\omega)] \\
  + \text{vtx.\ corr.}
\end{multline}

\subsection{Zero-density limit}
At zero density, the vertex corrections in \eqref{eq:chifreq} vanish
and the contact correlator is completely determined by the
particle-hole excitations of pairs \eqref{eq:zetaX}.  The pair
propagator (T matrix), in turn, is given diagrammatically by repeated
particle-particle scattering,
\begin{align}
  \label{eq:Tmatrix}
  \Gamma^R(\vec q,\varepsilon)^{-1}
  & = g_0^{-1} - \int \frac{d^dp}{(2\pi)^d} \,
    \frac{1-f(\xi_{\vec p})-f(\xi_{\vec q-\vec p})}
    {\varepsilon+i0-\xi_{\vec p}-\xi_{\vec q-\vec p}}.
\end{align}
Here, $f(\varepsilon)$ denotes the Fermi distribution and
$\xi_p=\varepsilon_p-\mu$ measures the dispersion
$\varepsilon_p=p^2/2m$ from the chemical potential $\mu$.  At zero
density there is only a single up and a single down fermion, such that
the Fermi functions $f(\xi_{\vec p})$ vanish: the momentum integral is
then performed analytically and yields the pair spectral functions
$A(\vec q,\varepsilon) = -(1/\pi)\Im\Gamma^R(\vec q,\varepsilon)$
given as
\begin{align}
  \label{eq:A2D}
  A_\text{2D}(\vec q,\varepsilon)
  & = \frac{4\pi}m \Bigl[\eb \delta(\varepsilon+2\mu+\eb-\omega_q) \\
  & \qquad\qquad + \frac{\Theta(\varepsilon+2\mu-\omega_q)}
    {\ln^2[(\varepsilon+2\mu-\omega_q)/\eb] + \pi^2}\Bigr], \notag\\
  \label{eq:A3D}
  A_\text{3D}(\vec q,\varepsilon)
  & = \frac{4\pi}{m^{3/2}}
    \Bigl[2\sqrt{\eb}\, \delta(\varepsilon+2\mu+\eb-\omega_q) \Theta(a) \\
  & \qquad\qquad + \frac1\pi\,
    \frac{\sqrt{\varepsilon+2\mu-\omega_q}\,\Theta(\varepsilon+2\mu-\omega_q)}
    {\varepsilon+2\mu+(ma^2)^{-1}-\omega_q}\Bigr].\notag
\end{align}
The first term is the bound-state peak, which appears always for
attractive interaction in 2D and for $a>0$ (BEC side) in 3D, followed
by the scattering continuum; $\omega_q=q^2/2M$ denotes the dispersion
of fermion pairs (molecular bound states) of mass $M=2m$.

In order to compute the bulk viscosity in the zero-density limit, one
has to set $T=0$ and chemical potential $\mu=-\eb/2$ at the threshold
of the two-body binding energy $\eb=\hbar^2/ma^2$ ($\mu=0$ when there
is no two-body bound state on the BCS side in 3D).  At zero density,
dissociating a bound state at $\varepsilon=\omega_q=0$ in 2D yields
Eq.~\eqref{eq:zeta02D} in the main text.  This satisfies the sum rule
\eqref{eq:sumrule},
$S_\text{2D,vac} = \mathcal C_0/4\pi m = \eb/V =
-(1/4)(\partial^2\mathcal E_0/\partial(\ln\abs a)^2)_s$ where
$\mathcal E_0 = -\mathcal C_0/4\pi m$.  In 3D, a vacuum bound state
exists only for $a>0$, and one finds Eq.~\eqref{eq:zeta03D}, which
again satisfies the sum rule \eqref{eq:sumrule} with
$S_\text{3D,vac} = \mathcal C_0/36\pi ma = -(1/9)(\partial^2\mathcal
E_0/\partial(\ln\abs a)^2)_s$ and $\mathcal E_0 = -\mathcal C_0/8\pi ma$.

\subsection{High-frequency limit}
The response at high frequencies is only sensitive to the behavior of
the contact correlation at short times, which factorizes in the limit
$\tau\to0$ as 
\begin{align*}
  \chi(\vec x,\tau)
  & = \vev{\Delta^\dagger \Delta(\vec x,\tau) \Delta^\dagger
  \Delta(0,0)} \\
  & \simeq \vev{\Delta(\vec x,\tau)\Delta^\dagger(0,0)}
  \vev{\Delta^\dagger \Delta(0,0)} \\
  & = m^2\Gamma(\vec x,\tau) \mathcal C(0,0).
\end{align*}
In this limit, the T matrix $\Gamma(\vec x,\tau)$ is unaffected
by finite density and approaches the zero-density form \eqref{eq:A2D},
\eqref{eq:A3D}.  Therefore, the high-frequency limit $\omega\to\infty$
of the bulk viscosity is proportional to the contact density and
decays with the asymptotic frequency dependence \eqref{eq:zetahigh}
quoted in the main text for $\omega\gg\eb$.

\subsection{Virial expansion}
The virial expansion to second order $\mathcal O(z^2)$ in the
fermionic fugacity $z=e^{\beta\mu}$ correctly describes the properties
of the interacting Fermi gas as long as the pair fugacity
$z_\text{pair} = z^2e^{\beta\eb} \ll1$ remains small.  Because vertex
corrections in Eq.~\eqref{eq:zetaX} appear only at higher order in
$z$, the second-order virial result is fully determined by the first
term in that equation.  Since the occupation factor $b(\varepsilon)$
is already order $z^2$, it suffices to use the zero-density form of
the pair spectral function \eqref{eq:A3D} and the Boltzmann
distribution to obtain
\begin{multline*}
  \frac{d^2}{c_d^2}\zeta(\omega)
  = \frac{\pi m^4}{\omega} \int \frac{d^3q}{(2\pi)^3} \int
  d\varepsilon\, A_\text{3D}(\vec q,\varepsilon) A_\text{3D}(\vec
  q,\varepsilon+\omega)\\
  \times z^2 e^{-\beta(\varepsilon+2\mu)} [1-e^{-\beta\omega}].
\end{multline*}
With the explicit form of $A_\text{3D}$ \eqref{eq:A3D} and in the new
frequency variable $y=\beta(\varepsilon+2\mu-\omega_q)$ one finds
\begin{multline*}
  \frac{d^2}{c_d^2}\zeta(\omega)
  = 16\pi^3mz^2\, \frac{1-e^{-\beta\omega}}{\omega}
    \int \frac{d^3q}{(2\pi)^3}\, e^{-\beta\omega_q} \int dy\, e^{-y} \\
    \times \Bigl[2\sqrt{\beta\eb} \delta(y+\beta\eb) \Theta(v)
    +\frac1\pi \frac{\sqrt y\,\Theta(y)}{y+v^2}\Bigr]\Bigl[ y\mapsto
    y+\beta\omega \Bigr],
\end{multline*}
where $v=(\lambda/a\sqrt{2\pi})$ denotes the dimensionless 3D
interaction parameter.  Since $\zeta(\omega)$ is even in frequency it
suffices to consider $\omega>0$; then there is the bound-continuum
contribution from the bound state $y=-\beta\eb$ in the first bracket
and the continuum $y+\beta\omega>0$ in the second.  On the other hand,
for $y>0$ one obtains the continuum-continuum contribution, and both
terms are combined to yield
\begin{multline*}
  \frac{d^2}{c_d^2}\zeta(\omega)
  = 32\sqrt2\pi^2mz^2\lambda^{-3}\, \frac{1-e^{-\beta\omega}}{\omega} \\
  \times \Bigl[
  2ve^{v^2}\Theta(v)\frac{\sqrt{\beta(\omega-\eb)}\,\Theta(\omega-\eb)}
  {\beta\omega} \\
  + \frac1\pi \int_0^\infty dy
  \frac{e^{-y}\sqrt{y(y+\beta\omega)}}{(y+v^2)(y+\beta\omega+v^2)}\Bigr].
\end{multline*}
Note that there appears no bound-bound contribution
$\delta(\omega)$ because an ideal Bose gas of pairs is scale
invariant.  Multiplication with the coefficient $(c_d/d)^2 =
T^2\lambda^2v^2/(288\pi^3)$ in 3D yields the dynamical bulk viscosity
\eqref{eq:zeta3Dvir}. 

Frequency integration determines the spectral weight of the terms in
brackets as
$S_\text{bc} = [2(1+2v^2)e^{v^2}\erf(v)-4v^2e^{v^2}+4v/\sqrt\pi]
\Theta(v)$ for the bound-continuum contribution, which becomes large
in the BEC limit.  Furthermore, the continuum-continuum contribution
has weight
$S_\text{cc} = [(1+2v^2)e^{v^2}(1-\erf(\abs v))-2\abs v/\sqrt\pi]$,
and by combining both one exhausts the sum rule \eqref{eq:sum3Dvir}.

The contact density is given to the same order in the virial expansion
by the occupied spectral function of pairs,
\begin{multline*}
  \mathcal C_\text{3D,vir}
  = m^2 \int\frac{d^3q}{(2\pi)^3} \int_{-\infty}^\infty d\varepsilon\,
  A(\vec q,\varepsilon) b(\varepsilon) \\
  = 16\pi z^2\lambda^{-4} \left[ 1+\sqrt\pi\,
    ve^{v^2}(1+\erf(v))\right]. 
\end{multline*}
This is equivalent to the derivative of the second virial coefficient,
$\mathcal C = 8\pi\sqrt{2\pi}z^2\lambda^{-4} \partial b_2(v)/\partial
v$ with $b_2(v) = -1/4\sqrt2 + (1/\sqrt2)(1+\erf v)e^{v^2}$,
$b_2' = \sqrt{2/\pi} + 2vb_2$, and $b_2'' = 2(b_2+vb_2')$.  The
contact grows monotonously with the interaction parameter $v$, as
shown in Fig.~\ref{fig:virdc}; this generalizes earlier results
\cite{yu2009, dusling2013}.

In order to compute the adiabatic derivative of the contact, one has
to keep the entropy per particle $s = \beta(h-\mu)$ fixed.  This is
given in terms of the enthalpy per particle $h=(p+\mathcal E)/n$ and
the chemical potential $\mu$, and at second order virial expansion one
finds
\begin{equation*}
  s = \frac52\,\frac{1+z(b_2-vb_2'/5)}{1+2zb_2}-\ln(z)
\end{equation*}
in extension of the Sackur-Tetrode entropy formula $s=5/2-\ln(z)$.
The adiabatic derivative with respect to $v$ is related to the grand
canonical derivative keeping $\mu$ and $T$ fixed, plus an additional
term adjusting $z'/z$ to compensate for the change in entropy:
$(\partial z^2b_2'/\partial v)_s = (2\frac{z'}{z}b_2'+b_2'')z^2$.  In
the BEC limit, $z'/z=-v$ and we find the adiabatic derivative
$(\partial z^2b_2'/\partial v)_s = 2z^2b_2$.  The corresponding
two-body contact is
$\mathcal C_0 = 4\pi n/a = 16\pi z^2\lambda^{-4} \sqrt\pi 2ve^{v^2}$,
and the adiabatic derivative
$(\partial \mathcal C_0/\partial v)_s = \mathcal C_0/v = 16\pi
z^2\lambda^{-4} \sqrt\pi 2e^{v^2}$ yields the bound-continuum
contribution to the sum rule \eqref{eq:sum3Dvir} \emph{without a
  bound-bound contribution}.  Similarly, the adiabatic derivative of
the continuum part of the contact at fixed fermionic entropy
$s=5/2-\ln(z)$ yields the continuum-continuum contribution to the sum
rule \eqref{eq:sum3Dvir}, thus confirming Eq.~\eqref{eq:sumrule} by
explicit computation in the high-temperature limit.

The dynamical bulk viscosity in \emph{two dimensions} is computed in a
completely analogous fashion to obtain the bulk viscosity quoted in
the main text.  Its total spectral weight is given by the sum rule
\eqref{eq:sumrule} with
\begin{multline*}
  S_\text{2D,vir}
  = 2z^2T\lambda^{-2} \Bigl[ \beta\eb e^{\beta\eb} \\
  + 2\int_0^\infty dy\,
  \frac{e^{-y}\ln(y/\beta\eb)}{[\ln^2(y/\beta\eb)+\pi^2]^2} \Bigr].
\end{multline*}
The sum rule agrees with the adiabatic derivative of the contact
density \cite{ngampruetikorn2013, barth2014}
\begin{multline*}
  \mathcal C_\text{2D,vir} = 16\pi^2z^2\lambda^{-4}\Bigl[ \beta\eb
  e^{\beta\eb} \\
  + \int_0^\infty dy\,
  \frac{e^{-y}}{\ln^2(y/\beta\eb)+\pi^2} \Bigr].
\end{multline*}

\section{Luttinger-Ward approach}
The Luttinger-Ward approach \cite{haussmann2007} to the attractive
Fermi gas is based on two-component fermions
$\psi_\sigma(\vec x,\tau)$ which interact by forming (virtual) pairs
$\Delta(\vec x,\tau)$.  It is constructed to be a conserving
approximation which exactly conserves not only particle number and
momentum current but also the dilatation current (scale invariance)
\cite{enss2011} and the Tan relations \cite{enss2012crit}.  The pair
propagator is given by the T matrix (cf.\ Eq.~\eqref{eq:Tmatrix})
\begin{align}
  \label{eq:Gamma}
  \Gamma^{-1}(\vec q,i\epsilon_n) = g_0^{-1} + M(\vec q,i\epsilon_n)
\end{align}
with bare coupling $g_0$, and the pair self-energy is given by the
fermion particle-particle bubble $M(\vec q,i\epsilon_n)$.  This is a
convolution of two \emph{dressed} fermion propagators
$G(\vec q,i\epsilon_n)$ in momentum $\vec q$ and Matsubara frequency
$i\epsilon_n$.  However, numerically it is more conveniently computed
by a Fourier transform to real space $\vec r$ and imaginary time
$\tau$, where the particle-particle bubble is a local product
\begin{align}
  \label{eq:M}
  M(\vec r,\tau) = [G(\vec r,\tau)]^2;
\end{align}
the resulting $M(\vec r,\tau)$ is then Fourier transformed back to
momentum and frequency $M(\vec q,i\epsilon_n)$ to be inserted in the
pair Dyson equation \eqref{eq:Gamma}.  In turn, the fermion propagator
for each spin component of the balanced gas is given by the Dyson
equation
\begin{align}
  \label{eq:G}
  G^{-1}(\vec q,i\epsilon_n)
  = -i\epsilon_n + \varepsilon_q - \mu + \Sigma(\vec q,i\epsilon_n)
\end{align}
with the fermionic self-energy $\Sigma(\vec q,i\epsilon_n)$ determined
by scattering fermions off virtual pairs.  The particle-hole bubble of
a pair (particle) and a fermion (hole) is again concisely written as a
local multiplication in Fourier space,
\begin{align}
  \label{eq:Sigma}
  \Sigma(\vec r,\tau) = \Gamma(\vec r,\tau) G(-\vec r,-\tau).
\end{align}
These four equations \eqref{eq:Gamma}--\eqref{eq:Sigma} form a closed
set of equations, which is solved self-consistently by iteration until
convergence is reached.  The propagators $\Gamma(\vec q,i\epsilon_n)$
and $G(\vec q,i\epsilon_n)$ are first initialized with the bare
propagators, then Fourier transformed to real space.  The
self-energies \eqref{eq:M} and \eqref{eq:Sigma} are computed in real
space, then Fourier transformed back to momentum space.  These are
then inserted into the Dyson equations \eqref{eq:Gamma} and
\eqref{eq:G}, which provide the starting point for the next iteration.

The Luttinger-Ward approach has previously been used for fermionic
shear and spin transport \cite{enss2011, enss2012spin}; by considering
the response to variations of an external field (shear strain or spin
gradient), a new set of self-consistent transport equations is
obtained for the renormalized current vertices which include vertex
corrections to satisfy the Ward identities exactly.  For the bosonic
contact correlations, instead, one has to consider the response to a
time-dependent scale variation.  Specifically, the scale breaking
variation $\delta=-\partial/\partial(4\pi ma)^{-1}$ at external drive
frequency $i\omega_m$ of each of the above equations
\eqref{eq:Gamma}--\eqref{eq:Sigma} yields a new set of self-consistent
transport equations for the renormalized trace anomaly response
vertices:
\begin{align}
  \label{eq:dGamma}
  \delta \Gamma^{-1}(\vec q,i\epsilon_n;i\omega_m) 
  & = \delta g_0^{-1} + \delta M(\vec q,i\epsilon_n;i\omega_m),\\
  \label{eq:dM}
  \delta M(\vec r,\tau;i\omega_m)
  & = 2G(\vec r,\tau;i\omega_m)\, \delta G(\vec r,\tau;i\omega_m),
  \\
  \label{eq:dG}
  \delta G^{-1}(\vec q,i\epsilon_n,i\omega_m) 
  & = \delta G_0^{-1}(\vec q,i\epsilon_n,i\omega_m) \notag\\
  & \quad + \delta \Sigma_\text{MT}(\vec q,i\epsilon_n,i\omega_m) \notag\\
  & \quad + \delta\Sigma_\text{AL}(\vec q,i\epsilon_n,i\omega_m), \\
  \label{eq:dSMT}
  \delta\Sigma_\text{MT}(\vec r,\tau;i\omega_m)
  & = \Gamma(\vec r,\tau;i\omega_m)\, \delta G(-\vec
    r,-\tau;i\omega_m),\\
  \label{eq:dSAL}
  \delta\Sigma_\text{AL}(\vec r,\tau;i\omega_m)
  & = \delta \Gamma(\vec r,\tau;i\omega_m)\, G(-\vec r,-\tau;i\omega_m).
\end{align}
In the first line \eqref{eq:dGamma}, $\delta \Gamma^{-1}$ denotes the
interaction scale variation of the bosonic pair propagator: it
consists of the bare contact vertex $\delta g_0^{-1} = -m^2$, which is
obtained from the scale variation $c_d$ of the bare coupling, and the
contact vertex correction $\delta M$, which arises from the scale
variation of the particle-particle bubble \eqref{eq:dM}.  While the
bare fermionic propagator has no scale dependence,
$\delta G_0^{-1}=0$, the dressed fermion propagator \eqref{eq:G}
acquires a scale dependence from the self-energy term
\eqref{eq:Sigma}.  This gives rise to a \emph{fermionic} trace anomaly
vertex $\delta G^{-1}$ in \eqref{eq:dG} with two distinct types of
vertex corrections.  The first, so-called Maki-Thompson vertex
correction $\delta\Sigma_\text{MT}$ in \eqref{eq:dSMT} is built from a
fermionic anomaly vertex $\delta G$, while the second,
Aslamazov-Larkin vertex corrections $\delta\Sigma_\text{AL}$ in
\eqref{eq:dSAL} arises from the renormalized bosonic contact vertex
$\delta\Gamma$.  The scale variation of the propagators is computed
from the inverse propagators as
\begin{align*}
  \delta G(\vec q,i\epsilon_n;i\omega_m)
  & = -G(\vec q,i\epsilon_n)\, \delta G^{-1}(\vec q,i\epsilon_n;i\omega_m)\\
  & \qquad \times G(\vec q,i\epsilon_n+i\omega_m), \\
  \delta\Gamma(\vec q,i\epsilon_n;i\omega_m)
  & = -\Gamma(\vec q,i\epsilon_n)\, \delta\Gamma^{-1}(\vec q,i\epsilon_n;i\omega_m)\\
  & \qquad \times \Gamma(\vec q,i\epsilon_n+i\omega_m).
\end{align*}
The self-consistent transport
equations are solved separately for each value of the external
frequency $i\omega_m$; as before, a Fourier transform converts the
anomaly vertices between Fourier space $(\vec q,i\epsilon_n)$ and real
space $(\vec r,\tau)$ at fixed parameter $i\omega_m$.

\begin{figure}[t]
  \centering
  \includegraphics[width=\linewidth]{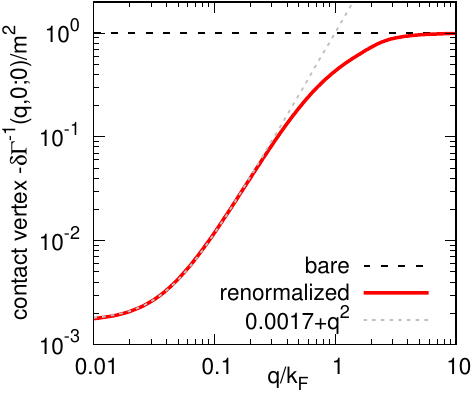}
  \caption{Renormalized contact vertex.  The static contact vertex
    near $T_c$ is strongly suppressed by contact vertex corrections as
    $\sim q^2$ at low momenta $q$.  Luttinger-Ward data are shown for
    reduced temperature $(T-T_c)/T_c=0.003$.}
  \label{fig:vertexcorr}
\end{figure}

In the first iteration of the transport equations, only the bare pair
propagator depends on scale and yields the bare contact vertex
$\delta g_0^{-1} = -m^2$, while the bare fermions are independent of
the interaction scale,
$\delta G_0^{-1}(\vec q,i\epsilon_n;i\omega_m) = 0$.  In subsequent
iterations, the fermionic anomaly vertex $\delta G^{-1}$ picks up an
interaction scale dependence from the MT and AL vertex corrections
\eqref{eq:dSMT} and \eqref{eq:dSAL}.  The scale dependence of the
fermionic vertex, in turn, generates vertex corrections $\delta M$
which renormalize the contact vertex $\delta\Gamma^{-1}$.  Once
convergence is reached, the Maki-Thompson and Aslamazov-Larkin vertex
corrections include contributions of arbitrarily high perturbative
order in the bare coupling.

The spatially integrated contact correlation function
\eqref{eq:chifreq} is computed by the Kubo formula
\begin{multline}
  \label{eq:XKubo}
  \chi(\vec k=0,i\omega_m) = X(i\omega_m)\\
  = \frac1{\beta V}\sum_{\vec q \epsilon_n} \delta g_0^{-1}\,
  \Gamma(\vec q,i\epsilon_n)\, \delta\Gamma^{-1}(\vec q,i\epsilon_n;i\omega_m)\,
  \Gamma(\vec q,i\epsilon_n+i\omega_m)
\end{multline}
for each value of the external Matsubara frequency $i\omega_m$.
Finally, analytical continuation to real frequencies $\omega+i0$ is
performed using Pad\'e approximants for the first few tens of
Matsubara frequencies to obtain the retarded contact correlator
$X^R(\omega)$ in \eqref{eq:Xret}, which yields the bulk viscosity via
Eq.~\eqref{eq:zetaX}.

At high temperature, the vertex corrections are subleading and it
suffices to use the bare contact vertex
$\delta \Gamma^{-1}=\delta g_0^{-1}$, such that \eqref{eq:XKubo}
simplifies to 
\begin{multline}
  \chi_\text{vir}(\vec k=0,i\omega_m) = X_\text{vir}(i\omega_m)\\
  = \frac{m^4}{\beta V}\sum_{\vec q \epsilon_n} 
  \Gamma(\vec q,i\epsilon_n)\, \Gamma(\vec q,i\epsilon_n+i\omega_m)
\end{multline}
in agreement with the first term in Eq.~\eqref{eq:chifreq}.  Hence,
the virial limit is contained within the LW approach.

In the BEC limit $a\to0^+$, the pair propagator has a large gap of
order $\eb\gg\ef,T$ between the bound state branch and the continuum
of dissociated pairs.  However, the bound state branch is no longer a
$\delta$ peak as in Eq.~\eqref{eq:A3D} but is broadened due to
atom-dimer and dimer-dimer scattering contained within the dressed
propagators and the vertex corrections in the LW equations.  This
broadening gives rise to a finite bound-bound contribution to the bulk
viscosity, which should approach the dc bulk viscosity of a weakly
repulsive Bose gas as $a\to0^+$.

\section{Importance of contact vertex corrections}

While in the high-temperature limit the fully dressed contact vertex
$\delta\Gamma^{-1}(\vec q,i\epsilon_n;i\omega_m)$ is close to the bare
contact vertex $\delta g_0^{-1}=-m^2$, it is found to be substantially
renormalized in the quantum degenerate regime approaching $T_c$.  In
particular, at the superfluid phase transition the pair propagator
$\Gamma\sim 1/(\xi^{-2}+q^2)$ becomes gapless and gives rise to
divergent critical fluctuations as the correlation length
$\xi\to\infty$ diverges.  One might expect that the contact
correlations \eqref{eq:XKubo} would also diverge with a positive power
of $\xi$.  However, the contact vertex corrections $\delta M$ which
are included within the Luttinger-Ward approach strongly suppress the
static contact vertex at small momenta and result in a scaling form
$\delta \Gamma^{-1}(\vec q,0;0) \sim q^2$, which renders the contact
correlation less singular when approaching $T_c$.  This is illustrated
in Fig.~\ref{fig:vertexcorr}, which shows the fully renormalized
contact vertex in units of the bare contact vertex $\delta g_0^{-1}$:
at large momenta $q\gg k_F$ it remains unrenormalized, but is strongly
suppressed as $\sim q^2$ for small momenta $\xi^{-1} \ll q\ll k_F$
before it eventually saturates for $q\ll\xi^{-1}$, where $\xi$ depends on
the distance from $T_c$.

\end{document}